\documentclass[prl,aps,showpacs,twocolumn]{revtex4}

\def\be{\begin{equation}}
\def\ee{\end{equation}}

\usepackage{graphicx}
\usepackage{braket}
\usepackage{bm}
\usepackage[latin1]{inputenc}

\begin{document}

\title{
Addressing Two-Level Systems Variably Coupled To An Oscillating Field}
\author{Nir Navon\footnote{Present email and address: nn270@cam.ac.uk, Cavendish Laboratory, University of Cambridge, J.J. Thomson Ave., Cambridge CB3 0HE, United Kingdom}, Shlomi Kotler, Nitzan Akerman, Yinnon Glickman, Ido Almog and Roee Ozeri}
\affiliation{Department of Physics of Complex Systems, Weizmann Institute of Science, Rehovot 76100, Israel}

\date{\today}

\begin{abstract}
We propose a simple method to spectrally resolve an array of identical two-level systems coupled to an inhomogeneous oscillating field. The addressing protocol uses a dressing field with a spatially-dependent coupling to the atoms. We validate this scheme experimentally by realizing single-spin addressing of a linear chain of trapped ions that are separated by $\sim3$ $\mu$m, dressed by a laser field that is resonant with the micromotion sideband of a narrow optical transition.

\end{abstract}

\pacs{37.10.De; 42.50.Ct; 67.85.-d; 03.67.Mn}

\maketitle

Single-qubit addressing is a pristine form of quantum control, and is an increasingly important tool to probe and manipulate small-scale quantum registers. It is a necessary requirement for quantum computing schemes \cite{blatt2012quantum}, while the use of single-spin imaging (and control) has recently allowed tremendous advances in the experimental study of quantum many-body systems \cite{bloch2012quantum}. One popular method relies on tighly focussed fields, so as to have non-negligible coupling only to the target particle \cite{naegerl1999laser,dumke2002micro}. This scheme usually requires an optical transition, and application of an optical field on diffraction-limited length scales (or below \cite{weitenberg2011single}), which is very challenging due to stability reasons.
Another general approach, inspired by magnetic resonance imaging, uses instead gradients of (electro-)magnetic field. By lifting the energy degeneracy through a position-dependent Zeeman- (or AC-Stark \cite{staanum2002trapped,haljan2005entanglement}) shift, single atoms can be addressed using the transition frequency that matches their location \cite{schrader2004neutral,wang2009individual,johanning2009individual,khromova2012designer,brahms2011cavity,lundblad2009field}.
These schemes require either magnetic-field sensitive states, or additional off-resonant levels to produce the AC-Stark shift. Yet another possibility arises if the addressing field can be engineered to have zero-coupling points,  or commensurate values in the case of two particles, as proposed and demonstrated with micromotion addressing in ion traps \cite{turchette1998deterministic,leibfried1999individual}.

\begin{figure}[h!]
\centerline{\includegraphics[width=1\columnwidth]{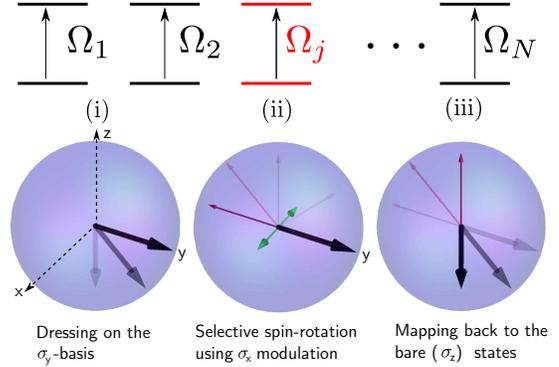}}
\caption{(Color online) The system considered is an array of $N$ identical non-interacting two-level systems, subjected to an external coupling field with a Rabi frequency
$\Omega_j$ for particle $j$. The lower panel shows a sketch of the single-spin addressing protocol, involving three steps: (i) the bare eigenstate of all particles $\uparrow$ and $\downarrow$ at the Bloch sphere poles are mapped onto the dressed eigenstates in the sphere equatorial plane (shown by thick black arrows). (ii) A driving field (shown by the green arrow) which is amplitude-modulated at a frequency $\Omega_j$ rotates the spin of particle $j$ (shown by the thin red arrow, and the red two-level system in the upper panel). (iii) The dressed states are mapped back on the bare basis at the sphere poles.
All particles end up in state $\downarrow$, except particle $j$ which is in $\uparrow$.
} \label{Fig1}
\end{figure}

In this letter, we propose a general scheme for selectively addressing an array of two-level systems by using a spatially-varying dressing field. The scheme has minimal requirements: it applies to perfect two-level systems (for which AC-Stark shift techniques are inefficient), and all frequencies necessary lie close to the transition frequency (contrary to ${\bf B}$-field gradients methods, which uses a DC component). In addition, the scheme is generic and equally works for optical, micro-wave or radiofrequency qubits, regardless of their magnetic-field sensitivity. Other systems could benefit from this scheme as various applications of dressed field control have been demonstrated for cold atomic gases \cite{gerbier2006resonant}, trapped ions \cite{timoney2011quantum}, NV centers \cite{belthangady2012dressed,cai2012robust} or superconducting qubits \cite{wilson2007coherence}. We experimentally demonstrate this protocol using a linear array of trapped ions. The inhomogeneous dressing field was provided in our setup by the inhomogenous micromotion of ions in the chain and a narrow linewidth laser that was tuned to the micromotion sideband of a narrow optical quadrupole transition.

We consider the Hamiltonian of the Rabi model, describing a non-interacting array of identical $N$ two-level systems inhomogeneously coupled to an external oscillating field. The Hamiltonian for particle $j$ reads
\begin{equation}
H_{\textrm{lab}}^{(j)}=-\frac{\hbar\omega_0}{2}
\sigma_z^{(j)}+
\hbar\Omega_j\cos(\omega_L t)
\sigma_y^{(j)},
\end{equation}
where $\omega_L$ is the coupling field frequency, and $\sigma_i^{(j)}$ ($i=x,y,z$) is the $i$-Pauli matrix acting on particle $j$. Without loss of generality, the phase of the field is arbitrarily set to realize the $\sigma_y$-coupling. Note that while the transition energy $\hbar\omega_0$ is assumed equal for all particles, the Rabi frequency associated with the coupling $\Omega_j$ is particle-dependent.
In the rotating-wave approximation (RWA), the Hamiltonian in the frame rotating at $\omega_L$ is
\begin{equation}\label{Hdress}
H_{\textrm{dress}}^0=\sum_j \frac{\hbar\Omega_j}{2} \sigma_y^{(j)}+\frac{\hbar\Delta}{2}\sigma_z^{(j)}=\sum_j E^{(j)}\sigma_{\phi_j}^{(j)},
\end{equation}
where the frequency detuning of the oscillating field to the atomic transition is $\Delta=\omega_0-\omega_L$, $E^{(j)}=\frac{\hbar}{2}\sqrt{\Omega_j^2+\Delta^2}$ and $\sigma_{\phi_j}^{(j)}=\sigma^{(j)}_y\cos\phi_j+\sigma^{(j)}_z\sin\phi_j$. The dressed states are the eigenstates of $\sigma_{\phi_j}^{(j)}$, with eigen-energies $\pm E^{(j)}$. It is thus possible to individually address particle $j$ by inducing spin-flips between the dressed states using a $\sigma_x$-perturbation in the rotating frame, resonant with the dressed states transition energy. While for far off-resonant dressing fields ($\Delta\gg\Omega_j$) the energy difference between neighboring two-level systems drops as $\sim\Omega_i/\Delta$, optimal spectral distinguishability is obtained for $\Delta=0$, i.e. for a resonant field with the transition. In that case, $H^0_{\textrm{dress}}$ is formally equivalent to the Hamiltonian of an array of spin-$1/2$ particles in a fictitious magnetic field pointing in the $y$-direction and having a gradient along the axis of the ion crystal. In other words, the two-level systems are coupled to a spatially varying dressing field, and the energies of the dressed states, which are the eigenstates of $\sigma_y^{(j)}$, are position-dependent. Spectroscopic resolution can thus be mapped onto spatial resolution.

This analogy enables to construct a simple protocol for individual spin rotations, as is illustrated in Fig.\ref{Fig1}: (i) The bare spin eigenstates ($\sigma_z$ basis) are mapped onto an eigenstate of the dressed spin ($\sigma_y$ basis)  $\ket{\uparrow}_d$/$\ket{\downarrow}_d$, and the strong resonant dressing field is turned on, locking the eigenstates in place by opening an energy gap between them. This energy gap is determined by the local coupling strength of each atom to the dressing field. (ii) A small $\sigma_x$ perturbation is amplitude-modulated (in the $\omega_0$-rotating frame), with a modulation frequency $\delta$, so that the Hamiltonian becomes $H_{\textrm{dress}}=H^0_{\textrm{dress}}+\sum_{j=1}^N\hbar\lambda_j\cos(\delta t)\sigma_x^{(j)}$. This Hamiltonian is solved for each particle $j$ by performing an additional change to a rotating frame, at an angular frequency $\Omega_j$, assuming $\lambda_j,|\delta-\Omega_j|\ll\delta$, and neglecting the counter-rotating terms in this frame of reference \cite{thimmel1999rotating,RWANote}. In that case, we readily find that the probability for spin $j$ to be $|\uparrow\rangle_d$, while all the others are $|\downarrow\rangle_d$ is
\begin{equation}\label{EqMultipleRabi}
P^{(j)}(t)=P_j(t)\prod_{l\neq j}(1-P_l(t)),
\end{equation}
where $P_i(t)=((\lambda_i/4\bar{\Omega}_i)^2 \sin^2(\bar{\Omega}_i t/2)$ is the Rabi precession, at a frequency $\bar{\Omega}_i=\sqrt{\Delta_i^2+\lambda_i^2/4}$ and the detuning in the $\Omega_i$-rotating frame is $\Delta_i=\Omega_i-\delta$. For all $\Omega_j$'s different from $\delta$, the spins remain locked on $y$-axis while the spin of particle $j$ is resonantly rotated around the $x$-axis. (iii) The dressed states are mapped back to the $\sigma_z$ bare basis. Following this procedure, all particles remain in their initial state except for particle $j$, whose spin has been rotated.

We experimentally demonstrated the above individual spin-addressing scheme using a chain of atomic ions, held in a linear Paul trap, and typically separated by $\sim3$ $\mu$m. The experimental setup is presented in detail elsewhere \cite{akerman2011quantum}. In short, we trapped and Doppler-cooled one or more $^{88}$Sr$^+$ ions in a linear Paul trap. We implemented a two-level system using the 5S$_{1/2,+1/2}$ state, hereafter referred to as the bright state, due its photon statistics during state-selective fluorescence detection \cite{keselman2011high}, and the long-lived 4D$_{5/2,+3/2}$ state, hereafter called dark state. These two states are connected by an optical, electric-quadrupole, transition at 674 nm. This transition is driven using a narrow-linewidth diode laser stabilized to an external high-finesse cavity \cite{keselman2011high}. In an ideal linear Paul trap, the micromotion, due to the ac-field modulated on the RF electrodes, vanishes along the symmetry axis of the trap.
However, rather than a line of nulled RF amplitude as in the ideal case, the boundary conditions imposed by the trap end-caps generate a linearly increasing RF amplitude around a null-amplitude point at the trap center \cite{berkeland1998minimization}. The inhomogenity of this excess micromotion along the chain is used as a spatially-dependent coupling of the ions to the laser field. When operating the laser on the micromotion sideband of the quadrupole transition, located at a detuning of $\Omega_{\textrm{RF}}=21.75$ MHz from the carrier transition, the resulting Rabi frequency depends on the ion micromotion amplitude via: $\Omega_j =\Omega_c J_1(\eta_j)$, where $\Omega_c$ is the carrier Rabi frequency, in the absence of modulation, $J_1$ is the first-order Bessel function of the first kind \cite{wineland1979laser}. The parameter $\eta_j={\bf k}\cdot {\bf x}_j$ is the micromotion Lamb-Dicke parameter of particle $j$ along the laser wavevector ${\bf k}$, and ${\bf x}_j$ is its micromotion amplitude. For micromotion amplitudes small compared with the transition wavelength, $\eta_j\ll 1$, the local Rabi frequency can be approximated by $\Omega_j\approx \eta_j.\Omega_c/2$. The micromotion amplitude of the ions in the chain can be controlled by applying a differential voltage to the trap end-caps and displacing the ions along the trap axis, or by varying the RF voltage on the trap electrodes.  \\

\begin{figure}[h!]
\centerline{\includegraphics[width=1\columnwidth]{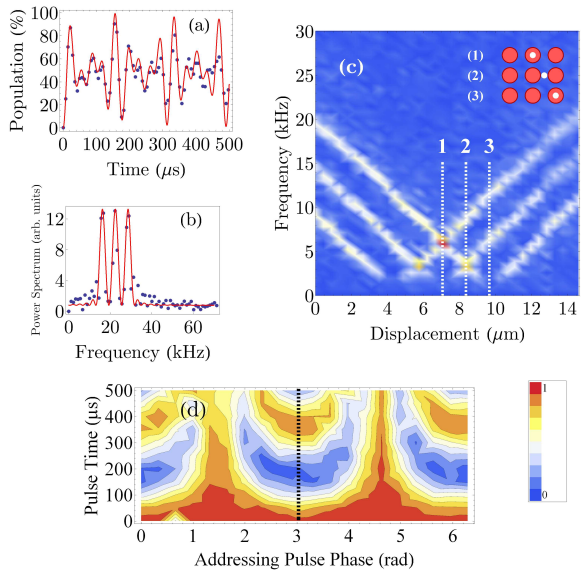}}
\caption{(Color online) Calibration of the single-spin addressing protocol. (a) and (b) show the time evolution, and Fourier spectrum respectively of the fraction of dark ions in the three-ion chain driven on the micromotion sideband, for a given axial position of the chain.
(c) Calibration of the dressed state energies by Fourier transform of the Rabi nutation curve on the micromotion sideband as a function of the axial displacement. The three particular points on the map (labelled 1 to 3) correspond to the RF-null being respectively on the central ion, in between two right-most ions and on the right-most ion, as shown in the schematic drawing on top of the map. The ions (RF-null) are represented by red (white) disks. (d) 2D scan of the phase of the amplitude-modulated driving field (with respect to the dressing beam phase) and the pulse time for a single ion. A $\sigma_x$ modulation is obtained for a phase $\phi\approx3.03$ rad (vertical dashed black line).}
 \label{Fig2}
\end{figure}

The values of the dressed state energies $\hbar\Omega_j$ for the different ions was first calibrated.
The strength of the dressing field was measured through the frequency of the Rabi oscillations on the micromotion sideband for different end-cap voltages. For multi-spin systems, the population of ions in the dark state was estimated from photon detection histograms. Fig.\ref{Fig2}a shows the fraction of ions in the dark state versus the micromotion sideband pulse time, for a three-ion chain. The different Rabi frequencies involved can be conveniently extracted by Fourier analysis of the Rabi nutation curve, as shown in Fig.\ref{Fig2}b. Three frequencies, differing by roughly 5 kHz are clearly observed. In Fig.\ref{Fig2}c, such spectra are shown as columns versus the global axial displacement of the chain. As seen, the three frequencies shift linearly with the displacement. This linearity proves negligible effect of the spatial inhomogeneity of the laser beam profile on the dressed state energies on the length scales of Fig.\ref{Fig2}. One frequency nulls every time one ion sits on the RF-null (points labeled 1 and 3 on Fig.\ref{Fig2}c).

In order to apply this scheme to a multi-spin system, one has to address simultaneously all spins for the steps (i) and (iii). This can be solved in two different ways. (1) If micromotion is
sufficiently weak so that the carrier Rabi frequency is almost constant throughout the chain, then
one can apply a collective carrier $\pi/2$-pulse (along $\sigma_x$) to bring the spins to the equatorial plane, to match the dressed states.
Here, the carrier and the micromotion sideband transitions, together with the RF signal for the trap electrodes must all be phase-locked.
(2) The spins are set on the equatorial plane using Rapid Adiabatic Passage (RAP) on the micromotion sideband interrupted on resonance. The RAP beam is kept on resonance, and thus automatically dresses the spins. In both schemes, following initialization, a AM-$\sigma_x$ pulse rotates the spin of the particle of interest. After the single-spin rotation the dressed states are mapped back to the bare states by reversing step (i).
While we have successfully implemented both methods, we preferred the former, since RAP is several times slower than the carrier $\pi/2$-pulses. \\

The amplitude-modulated $\sigma_x$ field was generated by adding a radio-frequency, that was amplitude-modulated at a frequency $\delta$, to the laser acousto-optic modulator (AOM). This signal had the same carrier frequency and is phase-shifted by $\pi/2$ from the dressing signal. Since the $\sigma_y$-dressing and the $\sigma_x$ driving field were provided by the same beam, then $\lambda_i=\alpha\Omega_i$, and the Rabi frequency $\bar{\Omega}_i$ is proportional to $\Omega_i$ when the detuning $\Delta_i$ vanishes. The phase of the driving field was tuned in the following way: after the spins are dressed, the phase and time of the modulation are scanned, while the driving frequency $\delta$ is tuned to a single-spin resonance. The resulting spin-flip probability for a single ion is displayed in Fig.\ref{Fig2}d. The period of the pattern is $\pi$ and we observed that at phases $\phi'$ and $\phi'+\pi$ (where $\phi'\approx1.46$) the scan is independent of time, corresponding to a modulation colinear with the dressing field ($\phi=\pm y$). At phases $\phi'\pm\pi/2$ (vertical dashed line in Fig.\ref{Fig2}d), we observe the highest contrast for the oscillations, corresponding to resonant $\sigma_x$ rotations of the dressed state. Note that the sum of the $\sigma_y$-dressing and the $\sigma_x$-modulation could be equally well produced by direct frequency modulation of the $\sigma_y$ beam.
\\

\begin{figure}[h!]
\centerline{\includegraphics[width=\columnwidth]{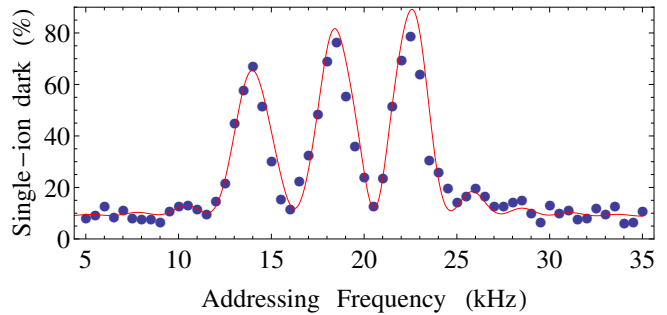}}
\caption{(Color online) Spectrum of the single-addressing protocol for a three-ion chain taken after a pulse time of 400 $\mu$s, in a trap whose secular frequencies were $(\omega_x,\omega_y,\omega_z)=2\pi\times(2.57,2.45,1.68)$ MHz. The RF frequency of the trap was 21.75 MHz.
} \label{FigSpectra}
\end{figure}

We next scanned the frequency of the driving field 400 $\mu$s pulse. A typical spectrum, displayed in Fig.\ref{FigSpectra}, exhibits three spectral features, corresponding to each ion addressed separately. The spectral splitting of the ions is given by the gradient of the coupling field, which in this case is proportional to the voltage applied on the RF electrodes. Using normal operation voltages, one reaches easily several kHz separation. This is an important figure of merit since small splittings require correspondingly longer pulses in order to frequency-resolve the peaks. Comparing to the method of individual addressing using magnetic field gradients \cite{johanning2009individual,wang2009individual}, the maximum splitting that we reached, of about 10 kHz, corresponds to a magnetic field gradient of about 50 G/cm (assuming a magnetic field sensitivity of 2.8 MHz/G), which is similar to the one obtained in \cite{johanning2009individual}. The Rabi frequency for each ion is easily deduced from a spectrum such as Fig.\ref{FigSpectra}, taken with a different pulse time.
For the 4 kHz splitting shown in Fig.\ref{FigSpectra}, a $\pi$-pulse of respectively 200, 280 and 560 $\mu$s leads to a spurious excitation of adjacent ions, which results from the overlap of their spectral response. The total crosstalk error \cite{crosstalkref} is estimated from Eq.(\ref{EqMultipleRabi}) to be respectively 1.6,1.1 and 4.1 $\%$ when addressing each ion respectively. The addressing fidelities of 90(2),88(2),85.6(2.5) $\%$ are consistent with an additional state preparation error per ion of $4$ $\%$, due to the spin-lock efficiency limited at long holding times. This is also consistent with the fidelity reached in a two-ion chain, of 94(2) $\%$ (with $\pi$-time of 180 $\mu$s). This limitation is not fundamental, and could be mitigated by increasing the power in the dressing field, or equivalently by increasing the RF voltage, both of which would increase the frequency gradient, and allow for shorter $\sigma_x$ operations. \\

\begin{figure}[h!]
\centerline{\includegraphics[width=0.6\columnwidth,height=0.5\columnwidth]{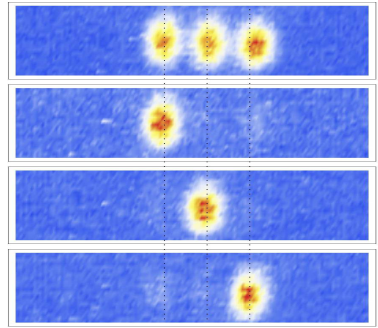}}
\caption{(Color online) Imaging the single-spin addressing protocol on a three-ion chain. The inter-ion spacing is 3 $\mu$m. Each image below is an average of 50 pictures taking on an EMCCD camera, with an exposure time of 20 ms. In the upper image, all three ions are in the S state, and therefore fluoresce. In the lower images a single ion remained in the S state while the other two were transferred to the D (dark) state. Because of the long exposure time required for the EMCCD imaging, there is a significant amount of error due to the finite lifetime of the dark state, in addition to the crosstalk errors. For instance on the second image, the error is respectively 7(4) $\%$ and 12(5) $\%$ for the central and rightmost ion respectively.
} \label{FigCamera}
\end{figure}

Lastly, we verify the spatially-selective nature of this protocol by taking images on an Electron Multiplying Charge Coupled Device (EMCCD) camera. In the upper panel of Fig.\ref{FigCamera}, we show an image of a three-ion chain, where all the ions are in the bright state. We then selectively flip each ion spin using the above protocol, and average over 50 images. We indeed observe successful single-spin addressing. The faint ``ghost'' images of adjacent ions are due to errors (see caption of Fig.\ref{FigCamera}). \\

To conclude, we have presented a method to single-qubit address an array of two-level systems, using a dressing field and working exclusively with frequencies close to the two-level transition. As such, it could be adapted to various quantum systems. It is possible to extend the above protocol to prepare arbitrary product state of single-qubit rotations. Indeed, while rotating spin $j$, particle $i$ precesses at a constant frequency $\Omega_i-\delta$ around the $y$-axis in the $\delta$-rotating frame. In order to leave all particles ($\neq j$) state unchanged, one can apply a $\sigma_x$-spin echo when reaching half the pulse time of spin $j$. This would refocus all particles spin to their initial value while preparing particle $j$ to its target state. With trapped ions, our protocol is well suited to address single ions in quantum registers using the typical intrinsic inhomogeneity of micromotion in RF Paul traps and without any additional elements or any need for optical resolution. This technique could be useful with continuous density distributions as well, to select atoms with respect to their coupling to some external field. As an example, it could be used in atomic fountain clocks to select atoms according to their coupling to a microwave cavity, where the inhomogeneities arise from the microwave spatial mode, and would provide an additional knob to study collisional or cavity phase shifts.

During the preparation of this manuscript, we became aware of a related work, realizing individual-ion addressing of a two-qubit chain using microwave field gradients on a microfabricated trap \cite{warring2012individual}.

We thank D. Leibfried and N. Davidson for helpful comments on the manuscript, and C. Salomon for fruitful discussions. This research was supported by the Israeli Science Foundation, the Minerva Foundation, the German-Israeli Foundation for Scientific Research, the Crown Photonics Center, the Wolfson Family Charitable Trust, Yeda-Sela Center for Basic Research, and David Dickstein of France.

\bibliographystyle{apsrev}

\begin{thebibliography}{21}
\bibitem[{RWA()}]{RWANote}
\bibinfo{note}{In the second RWA, all frequencies are of the same order of
  magnitude (in the kHz range), and one can thus expect some breakdown of the
  RWA.}
\bibitem[{crosstalkref()}]{crosstalkref}
\bibinfo{note}{Starting with all particles in $\ket{\downarrow}_d$, the crosstalk error when rotating the spin $j$ to $\ket{\uparrow}_d$ is $1-\textrm{Max}_t[P^{(j)}(t)]$. In other words, if any one particle state is incorrect, the result is an error regardless of whether the operation has been properly performed on the target qubit $j$.}
\end{thebibliography}

\end{document}